# Pressure effect on the melting temperature

Jozsef Garai[a] and Jiuhua Chen[a, b]

[a]Department of Mechanical and Materials Engineering, Florida International University, Miami, USA

[b]CeSMEC, Florida International University, Miami, USA

Pressure-melting temperature relationship is proposed and tested against the experiments of metals (Pt and Al), salt (NaCl), and ceramic (MgO) with positive results. The equation contains one open parameter which remains constant for the investigated substances. The constant value of the parameter indicates that the presented equation for the melting curve might be the first one which does not contain any arbitrary constant which is left open to fit to the experiments.

## 1. Introduction

The first observation of pressure effect on the melting point was made by Perkins in 1826. He compressed acetic acid [$CH_3COOH$] with 1100 Atm. pressure and found that the substance became crystallized [1]. Ever since then researchers try to describe the relationship between the pressure and the melting temperature from both theoretical and experimental aspects. Clapyeron described the pressure-temperature slope of the coexisting gas and liquid in 1834 which was extended to solid-liquid phase by Clausius [3]. The so called Clausius-Clapyeron relationship gives the first sound theoretical bases describing the pressure-temperature slope of the co-existing solid-liquid phase as:

$$\frac{dp}{dT} = \frac{\Delta H_f}{T \Delta V} = \frac{\Delta S}{\Delta V}. \tag{1}$$

where p is the pressure, T is the temperature, V is the volume, S is entropy and $H_f$ is the latent heat of fusion.

Assuming that the Lindemann melting criteria is correct and melting occurs when the amplitude of the atomic vibration reaches a certain fraction of the atomic diameter [4] gives the melting temperature as:

$$T_m = cMa^2\theta_D^2 \tag{2}$$

where c is a constant, M is the mass of the atom, $\theta_D$ is the Debye temperature, and a is the length parameter chosen to be the cube root of volume per atom as:

$$a = \sqrt[3]{\frac{V_{mol}}{N_A}} \tag{3}$$

where $N_A$ is the Avogadro's number and $V_{mol}$ the molar volume [5]. The melting temperature in Eq. (2) can also be expressed as the function of the molar volume and the Debye temperature as:

$$T_m = cV_m^{\frac{2}{3}}\theta_D^2. \tag{4}$$

Recently Zou and Chen [6] used the Lindenmann criteria for aluminum by using Chopelas-Boehler relationship [7] which assumes that the volume dependence of the Anderson-Gruneisen parameter is linear.

Simon and Glatzel [8, 9] suggested describing the pressure-melting temperature relationship as:

$$\frac{p_m - p_o}{a} = \left(\frac{T_m}{T_o}\right)^c - 1 \tag{5}$$

where $T_o$ and $p_o$ are coordinates of the triple point and a and c are constants characteristics of the substance. It has been shown by Kennedy [10] that Eq. (5) does not provide satisfactory basis when extrapolated to core pressure.

Kraut and Kennedy [11, 12] found linear relationship between the melting temperature and the isothermal decompression for many substances as:

$$T_m = T_m^0\left(1 + C\frac{\Delta V}{V}\right) \tag{6}$$

They claimed that in case of iron the linear relationship is valid up to compression of 0.5. It has been shown by Gilvarry [13] and Vaidya and Gopal [14] independently that using the Lindenmann law the constant in Eq. (6) can be expressed as a function of Gruneisen parameter as:

$$C = 2\left(\gamma - \frac{1}{3}\right) \tag{7}$$

It has also been shown [15, 16] that Eq. (6) can also be derived from the Claussius-Clapeyron relationship.

Wang et al. [17] proposed to introduce a critical temperature which is equivalent with the melting temperature. Using the EoS and conventional thermodynamic relationships the critical volume relating to this critical temperature is calculated. Deducting the thermal pressure from the actual pressure gives the pressure relating to the melting temperature.

Recently Kechin [18] proposed a melting curve in the form of

$$T_m = F(p)D(p). \tag{8}$$

where F(p) is the Simon equation for the rising part and D(p) is a dumping function. The equation is given as:

$$T_m = T_0\left(1 + \frac{\Delta p}{a}\right)^b e^{-c\Delta p}. \tag{9}$$

where a, b, and c are constants characteristics of the substance. Eq. (9) can successfully describe the maximum and negative slope of the melting curves of alkali metals.

New pressure-melting temperature relationship has been derived based on the assumption that the ratio of the phonon wavelength at the Debye and melting temperature remains constant [19]. The wavelength of the phonon vibration at the Debye temperature is equivalent with the smallest atomic unit of the crystal structure, which is the unit cell [20]. The physical description of melting suggests that at the melting temperature the thermal phonon vibration is in self-resonance with the lattice vibration of the surface atomic/molecular layer [21]. This self resonance occurs at a well defined temperature and triggers the detachment of the atomic/molecular sheet or platelets from the surface of the crystal [22]. Thus the wavelength of the phonon vibration at both the Debye and the melting temperature relates to lattice parameters as:

$$\overline{\lambda}_{T_D} \Rightarrow a/b/c \quad \text{and} \quad \overline{\lambda}_{T_m} \Rightarrow \frac{1}{n}d \quad \text{where } n \in \mathbb{N}. \tag{10}$$

If pressure is applied then the lattice parameters change their size in proportion and the ratio of the two lengths related to the lattice should remain constant.

$$\frac{\overline{\lambda}_{T_D}}{\overline{\lambda}_{T_m}} = \frac{\overline{f}_{T_m}}{\overline{f}_{T_D}} = \text{const}. \tag{11}$$

Knowing the boundary conditions and using Eq. (11) allows calculating the melting temperature for any pressure. At temperatures higher than the Debye temperature the relationship between the phonon frequency and the temperature is given as:

$$hf = k_B T \tag{12}$$

where h is the Planck constant, f is the frequency, $k_B$ is the Boltzmann constant and T is the temperature. The temperature at melting then can be expressed as function of the Debye temperature as:

$$T_m = \frac{\overline{f}_{T_m}}{\overline{f}_{T_D}} T_D = \frac{\overline{\lambda}_{T_D}}{\overline{\lambda}_{T_m}} T_D. \tag{13}$$

where $\overline{f}_{T_D}$ and $\overline{f}_{T_m}$ are the average frequency and $\overline{\lambda}_{T_D}$ and $\overline{\lambda}_{T_m}$ are the average wavelength of the phonon vibration at the Debye and melting temperature respectively. It has been suggested that the independent vibration of the atoms at temperatures higher than the Debye temperature results in vibration related electron shell deformation which changes the volume [19]. This effect can be positive or negative depending on the strength of the electron shell. Weak shell results in volume decrease because the deformations of the shells "absorb" the volume while the strong shell corresponds to volume increase. The correspondent thermodynamic parameter describing the volume change resulting from vibration related electron shell deformation is the temperature derivative of volume coefficient of thermal expansion. Applying pressure on a substance results in not only elastic deformation but also in volume change relating to the vibration related electron shell deformation. In order to take into consideration this effect a pressure term is introduced and Eq. (13) is modified as:

$$T_m = \left(\frac{\bar{f}_{T_m}}{\bar{f}_{T_D}} + c_{vib-vol}\frac{\alpha_1}{K_o}p\right) T_D. \qquad (14)$$

where $c_{vib-vol}$ is constant. Eq. (14) has been tested against the melting curve of Na with positive result [19]. In this study substances with positive temperature derivative of volume coefficient of thermal expansion [$\alpha_1$] are selected. Using the experimental melting curves of two metals, Al and Pt, one salt, NaCl and one ceramic MgO, the constants relating to the vibration related electron shell deformation are determined.

## 2. Method of calculation

The frequencies of the phonon vibrations can be calculated as:

$$f = \frac{v_B}{\lambda} \qquad (15)$$

where $v_B$ is the bulk sound velocity which is approximated as:

$$v_B = \sqrt{\frac{K}{\rho}} = \sqrt{\frac{KV_{mol}}{M}} \qquad (16)$$

K is the bulk modulus, $\rho$ is the density, M is the molar mass and $V_{mol}$ is the molar volume. The molar volume can be directly calculated by using the EoS of Garai [23] as:

$$V_{mol} = V_{o,mol} e^{\frac{-p}{ap+bp^2+K_o} + (\alpha_o+cp+dp^2)T_{m,p} + \left(1+\frac{cp+dp^2}{\alpha_o}\right)^f gT_{m,p}^2} \qquad (17)$$

where subscript o refers to the initial value at zero pressure and temperature. Thus $V_{o,mol}$ is the initial molar volume, $K_o$ is the initial bulk modulus and $\alpha_o$ is the initial volume coefficient of thermal expansion. In Eq. (17) a is a linear, b is a quadratic term for the pressure dependence of the bulk modulus, c is linear and d is a quadratic term for the pressure dependence of the volume

- 5 -

coefficient of thermal expansion and f and g are parameters describing the temperature dependence of the volume coefficient of thermal expansion. The theoretical explanations for Eq. (17) and the physics of the parameters are discussed in detail [23].

Assuming that the product of volume coefficient of thermal expansion and the bulk modulus is constant allows calculating the bulk modulus [24] as:

$$K = \left(K_o + K_o' p\right) e^{-(\alpha_o + \alpha_1 T)\delta T} \tag{18}$$

where $K_o'$ is the pressure derivative of the bulk modulus, $\alpha_1$ is a linear term expressing the temperature effect on the volume coefficient of thermal expansion and $\delta$ is the Anderson-Grüneisen parameter, which defined as:

$$\delta_T \equiv \left(\frac{\partial \ln B_T}{\partial \ln V}\right)_p = -\frac{1}{\alpha_{V_p}}\left(\frac{\partial \ln B_T}{\partial T}\right)_p = -\frac{1}{\alpha_{V_p} B_T}\left(\frac{\partial B_T}{\partial T}\right)_p. \tag{19}$$

The parameters in Eq. (18) are determined by fitting the universal p-V-T form of the Birch-Murnaghan EoS against experiments. The EoS is given as [25]:

$$p = \frac{3K_o e^{-(\alpha_o+\alpha_1 T)\delta T}}{2}\left[\left(\frac{V_o e^{(\alpha_o+\alpha_1 T)T}}{V}\right)^{\frac{7}{3}} - \left(\frac{V_o e^{(\alpha_o+\alpha_1 T)T}}{V}\right)^{\frac{5}{3}}\right]$$
$$\left\{1 + \frac{3}{4}\left(K_o' - 4\right)\left[\left(\frac{V_o e^{(\alpha_o+\alpha_1 T)T}}{V}\right)^{\frac{2}{3}} - 1\right]\right\} \tag{20}$$

The average wavelength of the phonon vibration relating to the Debye temperature is approximated as:

$$\bar{\lambda}_{TD} = a; c = \sqrt[3]{\frac{n_a V_{mol}}{N_A}} \tag{21}$$

where $n_a$ is the number of atoms in the unit cell and $N_A$ is the Avogadro's number. Using Eq. (12) and substituting Eqs. (15), (16) and (21) allows calculating the Debye temperature as:

$$T_D = hk_B^{-1} M^{-\frac{1}{2}} N_A^{\frac{1}{3}} n_a^{-\frac{1}{3}} V(p,T)_{mol}^{\frac{1}{6}} K(p,T)^{\frac{1}{2}} \tag{22}$$

Substituting Eqs. (17) and (18) into Eq. (22) allows calculating the Debye temperature at the given pressure and the melting temperature as:

$$T_{D(T_m,p)} = \frac{h}{k_B \sqrt{M}} \sqrt[3]{\frac{N_A}{n_a}} \left( nV_o^m e^{\frac{-p}{ap+bp^2+K_o} + \left(\alpha_o + cp + dp^2\right)T_{m,p} + \left(1 + \frac{cp+dp^2}{\alpha_o}\right)^f gT_{m,p}^2} \right)^{\left(\frac{1}{6}\right)} \times$$

$$\sqrt{(K_o + K_o'p)} \, e^{-(\alpha_o + \alpha_1 T_{m,p})\delta T_{m,p}} \qquad (23)$$

The melting temperature can be calculated by repeated substitution of Eq. (14) as:

$$T_{m,p} = \lim_{n \to \infty} f^n(T_{m,p}) \qquad (24)$$

where

$$f^n(T_{m,p}) = \left( \frac{T_{m0}}{T_{D0}} + c_{vib-vol} \frac{\alpha_1 p}{K_o} \right) \frac{h}{k_B \sqrt{M}} \sqrt[3]{\frac{N_A}{n_a}} \times$$

$$\left( nV_o^m e^{\frac{-p}{ap+bp^2+K_o} + \left(\alpha_o + cp + dp^2\right)T_{m,p(n-1)} + \left(1 + \frac{cp+dp^2}{\alpha_o}\right)^f gT_{m,p(n-1)}^2} \right)^{\left(\frac{1}{6}\right)} \times$$

$$\sqrt{(K_o + K_o'p)} \, e^{-(\alpha_o + \alpha_1 T_{m,p(n-1)})\delta T_{m,p(n-1)}} \qquad (25)$$

$n \in \mathbb{N}^*$ and $T_{m(0)} = T_{m0}$

where $T_{m0}$ is the melting temperature of the substance at atmospheric pressure and $T_{D0}$ is the Debye temperature at atmospheric pressure and the melting temperature. The convergence is fast and 15-20 iteration is sufficient.

## 3. Results and discussion

The melting curves of the investigated substances, Al [26-28], Pt [29], NaCl [30-32], and MgO [33-34] have been collected from the literature. The high pressure and temperature experiments of Al [35-37], Pt [36-39], NaCl [40] and MgO [25 and ref. therein] are used to determine the EoSs. The parameters of the B-M and the G EoSs were determined by unrestricted fitting. The results are listed in Table 1 and 2.

In the EoS of Garai [Eqs. (17)] symbol g represents the temperature derivative of the volume coefficient of thermal expansion. The fitting against experimental melting curves have been tested by using both g and $\alpha_1$. The value of the open parameter [$c_{vib\text{-}vol}$] in Eq. (25) has been determined for each of substances. The calculated $c_{vib\text{-}vol}$ values are very close to each other indicating that this parameter might be universal. The pressure term in Eq. (14) becomes dominant at high pressures if the ratio of g and the initial bulk modulus is high (ex. Al and NaCl). No theory is offered here to explain this behavior. Using $c_{vib\text{-}vol} = 2.5 \times 10^8 \, K^2$ and the g values the calculated meting curves are plotted against experiments (Fig. 1). The agreements with the experimental values are excellent.

## 4. Conclusions

The recently derived pressure melting temperature relationship has been tested against the experimental melting curves of metals, Al, Pt, salt, NaCl and ceramic MgO. The thermodynamic parameters required for the calculations have been determined from previous experiments by unrestricted fitting. The melting curve expression contains one open parameter which has been determined for the four melting curves. It is concluded that one single value [$2.5 \times 10^8 \, K^2$] is sufficient to reproduce the experimental values indicating that the given melting relationship might be universally applicable to substances with positive g or $\alpha_1$ parameters.

**Acknowledgement**

This research was supported by NSF Grant #0711321.

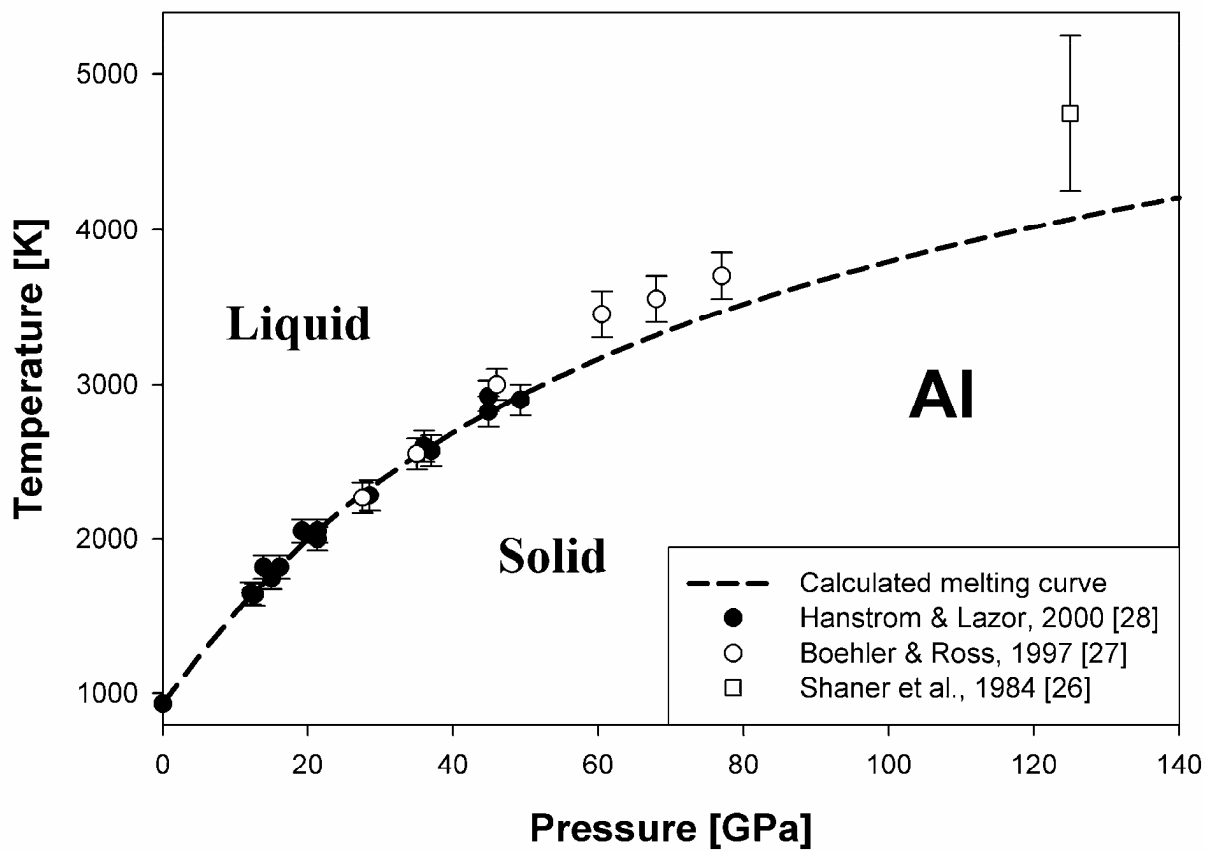
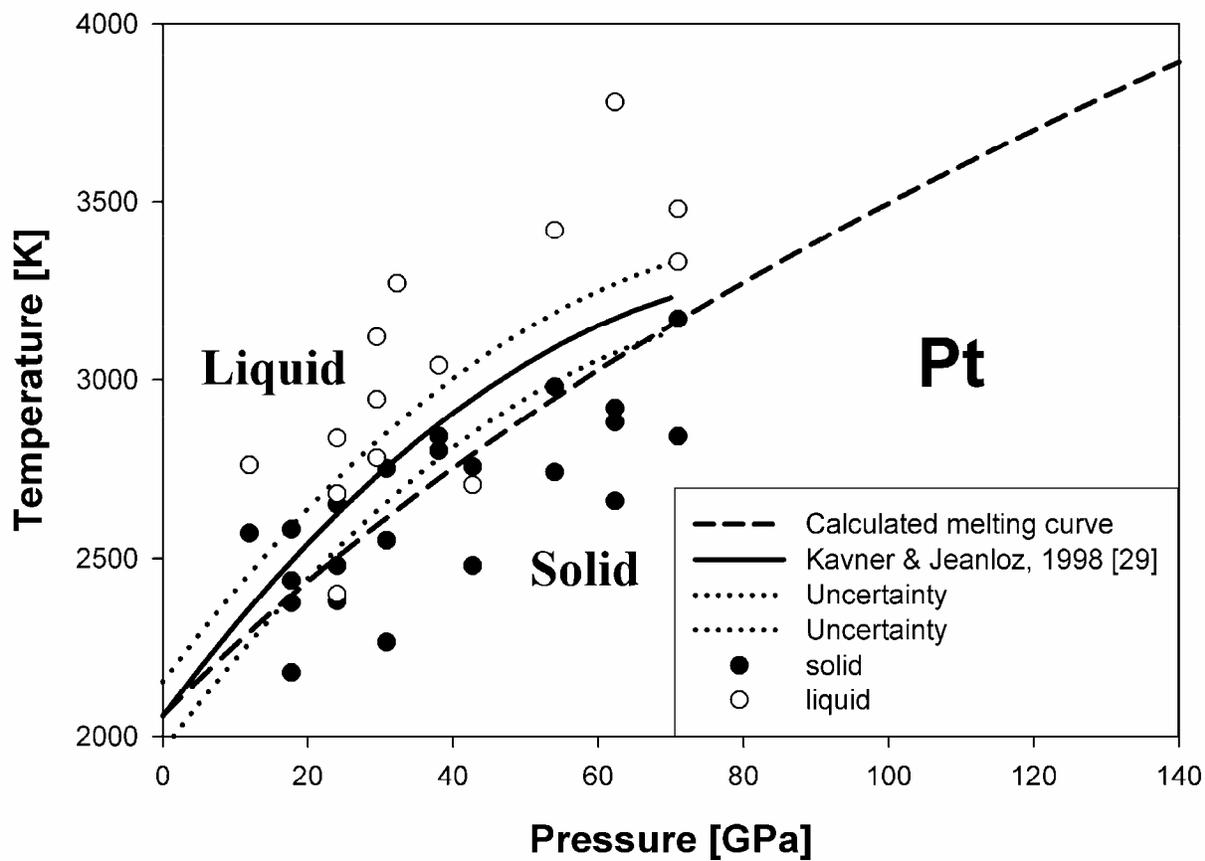

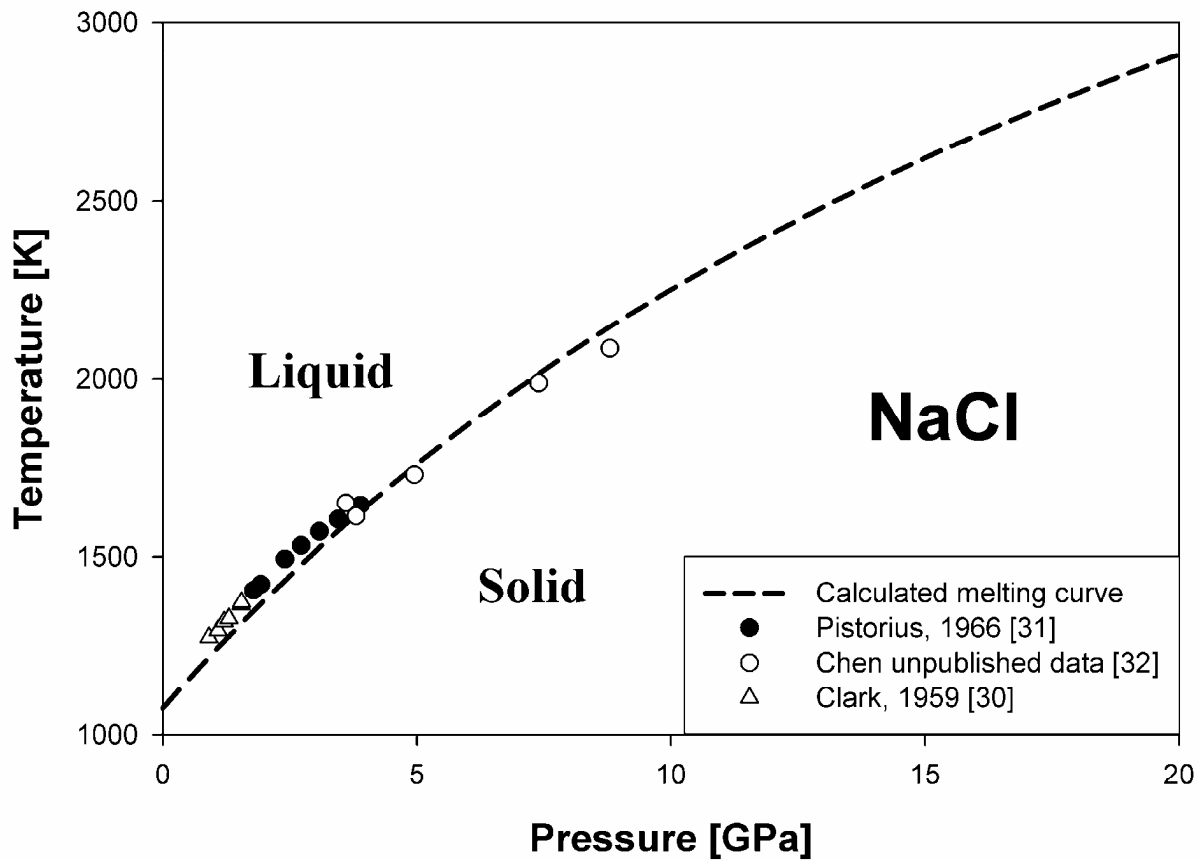
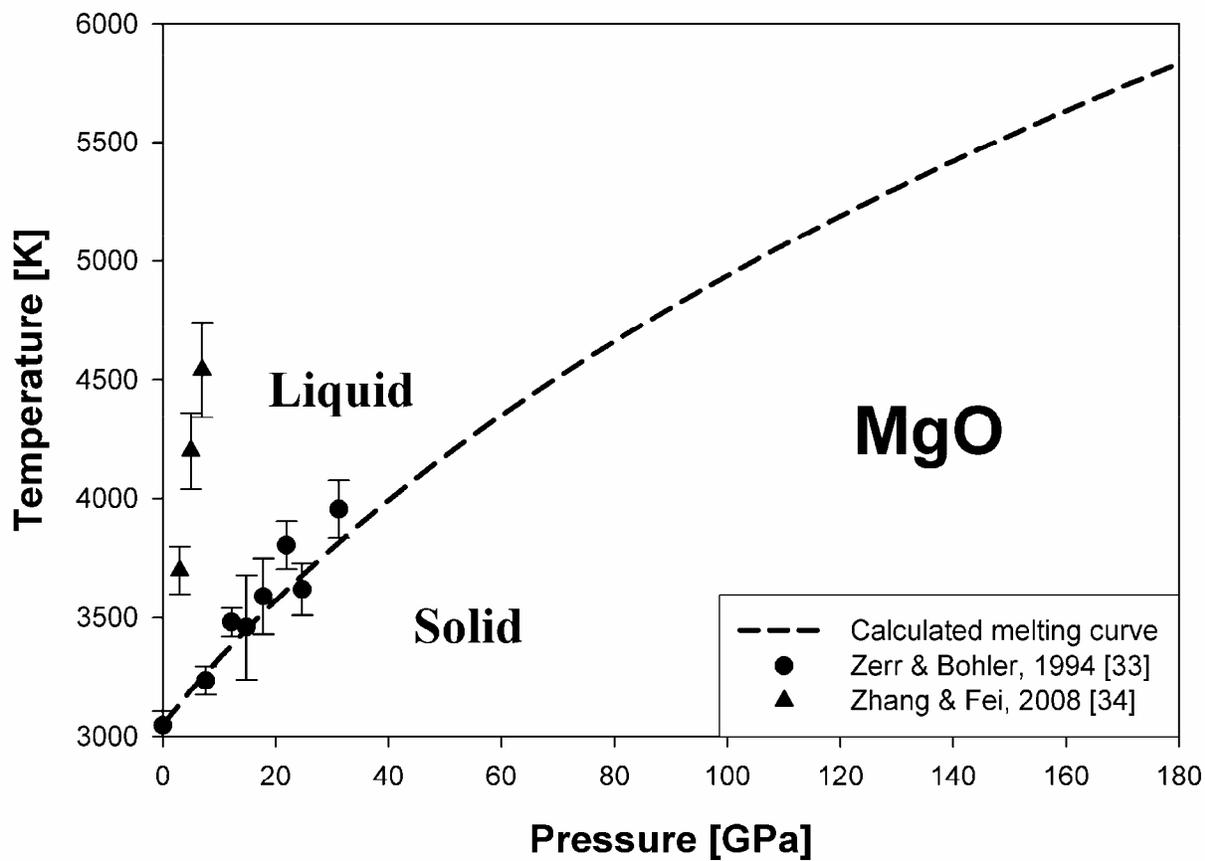

**Fig. 1.** The calculated melting curves for Al, Pt, NaCl and MgO are plotted against experiments. The melting curve of MgO [34] is extrapolated from FeO-MgO phase diagram.

**Table 1**. Parameters of the Birch-Murnaghan EoS.

| EoS-B-M | $K_o$ [GPa] | $K_o'$ | $\alpha_o$ x$10^{-5}$K$^{-1}$ | $\alpha_1$ x$10^{-9}$ K$^{-2}$ | $\delta$ | $\overline{\lambda_{T_D}}/\overline{\lambda_{T_m}}$ |
|---|---|---|---|---|---|---|
| MgO | 159.6 | 4.219 | 2.888 | 0.68 | 3.556 | 5.48 |
| Pt | 254.65 | 6.03 | 1.75 | 5.52 | 4.737 | 5.65 |
| Al | 84.93 | 3.79 | 4.7 | 34.15 | 4.7 | 1.68 |
| NaCl | 26.211 | 4.888 | 8.21 | 43.324 | 4.061 | 4.79 |

**Table 2**. Parameters of the Garai EoS.

| EoS-G | $V_o$ [cm$^3$] | $K_o$ [GPa] | $\alpha_o$ [$\times 10^{-5}$K$^{-1}$] | a | b [$\times 10^{-3}$] | c [$\times 10^{-7}$Pa$^{-1}$K$^{-1}$] | d [$\times 10^{-10}$Pa$^{-2}$K$^{-1}$] | g [$\times 10^{-9}$K$^{-2}$] | f |
|---|---|---|---|---|---|---|---|---|---|
| MgO | 11.142 | 165.15 | 2.957 | 1.721 | -2.249 | -2.090 | 4.4 | 6.903 | 10.3 |
| Pt | 9.041 | 282.42 | 1.554 | 2.234 | -3.375 | -0.283 | 0 | 8.631 | 38.2 |
| Al | 9.807 | 82.89 | 4.57 | 1.456 | -2.426 | -0.423 | 0 | 35.38 | 12 |
| NaCl | 26.323 | 26.459 | 6.67 | 1.964 | -16.744 | -20.562 | 0 | 61.714 | 10.7 |